\documentclass[aps,showkeys,preprint,floatfix]{revtex4}
\usepackage{graphicx}
\usepackage{graphics}
\usepackage{epsfig}
\usepackage{psfig}
\usepackage{bm}
\def\ie{i.e., }
\def\etal{{\it\ et.al.}}
\def\pr{\prime}
\def\be{\begin{equation}}
\def\ee{\end{equation}}
\def\bea{\begin{eqnarray}}
\def\eea{\end{eqnarray}}
\begin{document}
\title{EFFECTS OF MAGNETIC FLUX AND OF ELECTRON\\
       MOMENTUM ON THE TRANSMISSION\\AMPLITUDE IN THE AHARONOV-BOHM INTERFEROMETER
       \footnote{This work forms a part of the unpublished project report of the first author
                 submitted for partial fulfillment of the degree of Master
                 of Science in the department of Physics of Sri Sathya Sai Institute of
                 Higher Learning (Deemed University), Vidya Giri, Prasanthi Nilayam-
                 516\ ~134, A.P., India.}
  }
\author{M.V. AMARESH KUMAR}
\affiliation{Sri Sathya Sai Institute of Higher Learning, Vidhya
Giri, Prashanthi Nilayam-515 134, A.P., India.}
\email{amar@rri,res.in} \altaffiliation[Present address: ]{Raman
Research Institute, C.V.
Raman Avenue, Sadashivanagar, Bangalore 560 080, India.\\
}
\author{DEBENDRANATH SAHOO}
\affiliation{Institute of Physics, P.O. Sainik School, Bhubaneswar
751 005, Orissa, India} \email{dsahoo@iopb.res.in}
\altaffiliation[Permanent affiliation during the period of work:
]{Materials Science Division, Indira Gandhi Centre for Atomic
Research, Kalpakkam, Tamil Nadu-603 102, India. (presently
Visiting Professor, Sri Sathya Sai Institute of Higher Learning,
Vidhya Giri, Prashanti Nilayam 515134, Andhra Pradesh, India.)}
\date{}

\begin{abstract}

A characterization of the two-terminal open-ring Aharonov-Bohm
interferometer is made by analyzing the phase space plots in the
complex transmission amplitude plane. Two types of plots are
considered: type I plot which uses the magnetic flux as the
variable parameter and type II plot which uses the electron
momentum as the variable parameter. In type I plot, the trajectory
closes upon itself only when the ratio $R$ of the arm lengths (of
the interferometer) is a rational fraction, the shape and the type
of the generated flower-like pattern is sensitive to the electron
momentum. For momenta corresponding to discrete eigenstates of the
perfect ring (\ie the ring without the leads), the trajectory
passes through the origin a certain fixed number of times before
closing upon itself, whereas for arbitrary momenta it never passes
through the origin. Although the transmission coefficient is
periodic in the flux with the elementary flux quantum as the basic
period, the phenomenon of electron transmission is shown not to be
so when analyzed via the present technique. The periodicity is
seen to spread over several flux units whenever $R$ is a rational
fraction whereas there is absolutely no periodicity present when
$R$ is an irrational number. In type II plot, closed trajectories
passing through the origin a number of times are seen for $R$
being a rational fraction. The case $R=1$ (\ie a symmetric ring)
with zero flux is rather pathological--it presents a closed loop
surrounding the origin. For irrational $R$ values, the
trajectories never close.

\end{abstract}
\pacs{} \keywords{Aharonov-Bohm effect; transmission amplitude;
transmittance}
\maketitle
\section{Introduction}
\begin{figure}[h]
  \includegraphics[width=2in]{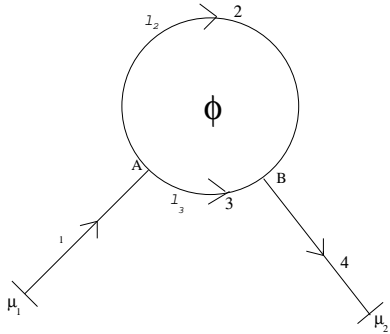}   
\caption{\label{ABI}Schematic diagram of the Aharonov-Bohm interferometer}
\end{figure}

The Aharonov-Bohm interferometer (ABI) in a two-terminal
configuration has come to stay as a reliable tool in the study of
mesoscopic/ nanoscopic systems. Figure\ref{ABI} presents a schematic
diagram of the ABI.
 The phase of the electron wave function has a
topological contribution arising due to the magnetic flux confined
within the ring, threading perpendicular to the ring plane. It also
has a dynamic contribution arising due to the electron momentum. In a
simplistic view $t$ is often assumed to be of the form
$$
t = t_1 + t_2 e^{i\phi}
$$
where $\phi$ is the Aharonov-Bohm (AB) phase and $t_1$, $t_2$ are
the TA's through the two arms of the ring. Whereas this view is
appropriate to the original free-space double-slit setup, it is not
so to the two-leaded ABI. More complicated dependence on $\phi$ as
well as the dynamical phases $\chi_1$ and $\chi_2$ contributed by
the two arms dictate the behavior of the TA\cite{butt-im-azbel,
cahay}. It is the purpose of the present paper to present our
analysis of the detailed effects of these phases on the TA.
Chen\etal\cite{chen_shi} report the effect of $\phi$ on
anti-resonances of the transmittance $T(={\mid t\mid}^2$ in the ABI.
They correlate the nature of constant-$T$ contour plots in the
complex electron-energy plane to the variation of $T$ with the
electron energy. Kim\etal\cite{kim_cho_kim_ryu} study the effects of
broken time-reversal symmetry on the transmission zeros of the ABI.
They propose a classification scheme of the phase trajectories of
$t$ in the complex $t$-plane and study the effect of the variation
of the $\chi$'s. Taniguchi and Buttiker\cite{tanibutt}, in a
pioneering work, analyze the Friedel phases and the TA phases for a
one-dimensional resonant tunnel double barrier, a wire with a side
branch and also for the combination of the above two structures.

In this work we consider the transmission properties of the
two-leaded AB ring (see Fig.{\ref{ABI}). We have repeated the
calculation of Xia\cite{xia} for the ABI for obtaining the TA. The
model and the assumptions that we make are same as the ones made
in his work\cite{xia}. The wave functions on the various segments
are assumed to be of the following form:
\begin{eqnarray}
\psi_1(x_1) &=& e^{ikx_1}+ae^{-ikx_1},\nonumber\\
\psi_2(x_2) &=& c_1e^{ik_1x_2}+d_1e^{-ik_2x_2},\nonumber\\
\psi_3(x_3) &=& c_2e^{ik_2x_3}+d_2e^{-ik_1x_3},\nonumber\\
\psi_4(x_4) &=& te^{ikx_4}.
\label{eq:wave functions}
\end{eqnarray}
The subscripts on the $\psi$'s correspond to the labels used on
the segments (see Fig.\ref{ABI}). The lengths of the segments
marked $2$ and $3$ are $l_2$ and $l_3$ and the total circumference
of the ring is $l=l_2+l_3$. Here  $x_1$, $x_2$ and $x_3$ have
common origin at A and $x_4$ has its origin at B. The boundary
conditions used at the junctions are
\begin{eqnarray}
&&\psi_1(0)=\psi_2(0)=\psi_3(0),\nonumber\\
&&\psi_2(l_2)=\psi_3(l_3)=\psi_4(0),\nonumber\\
&&\psi_1^\pr(0)=\psi_2^\pr(0)+\psi_3^\pr(0),\nonumber\\
&&\psi_2^\pr(l_2)+\psi_3^\pr(l_3)=\psi_4^\pr(0).\nonumber\\
\label{eq:boundary conditions}
\end{eqnarray}
Here $k_1=k+\eta$ and $k_2=k-\eta$ and $\eta = (2\pi/l)(\Phi
/\Phi_0)=(2\pi /l)\theta$, $\Phi$ denotes the magnetic flux
threading the ring and $\Phi_0=hc/e$ is the flux quantum. It is convenient
to use a dimensionless wave number $q$ defined by $k=(2\pi /l)q$ and
$\theta$, the dimensionless flux.
We have used the software MAPLE to obtain analytic expression for the TA.

We report on the characteristic features of the complex $t$-plane plots
when the magnetic flux is made to vary over one or several flux periods.
We refer to such plots as type I $t$-plot. Next we examine similar
plots when the electron momentum is varied. We refer to such plots as
type II plots.

\section{Type I $t$-plots}
We have analyzed the complex $t$-plane plots for a symmetric (\ie
$l_2=l_3$) as well as for an asymmetric (\ie $l_2\ne l_3$) ring.
We report our findings under separate headings for these two
cases. Note that our usage of the phrases ``symmetric'' and
``asymmetric'' ABI is different from the conventional usage.
\subsection{The symmetric ring interferometer}
For a symmetric ring ABI, for arbitrary values of $q$, the typical
behavior of $T$ as $\theta$ is varied is shown in the curves
marked $1$ and $2$ in Fig.\ref{transones}.
\begin{figure}[h]   
\vskip 3cm
  \includegraphics[width=2in]{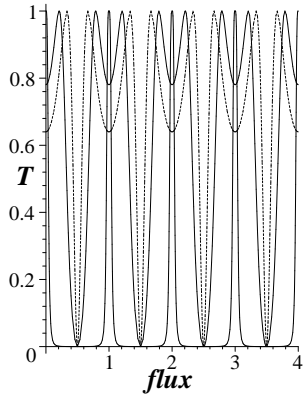}
\caption{\label{transones}Variation of $T$ with $\theta$. Thick
  line: $q=1$, thionine: $q=1/4$ and dotted line: $q=1/2$.}
\end{figure}
Note the presence of minima of $T$ at integer $q$ values. In
contrast, sharp transmission ones with characteristic appearance
of resonances arise for values of $q$ in close proximity of
integer $q$'s corresponding to discrete energy eigenstates. These
resonances were missed in the calculation of Xia\cite{xia}, but
their existence has been anticipated by Cahay \etal\cite{cahay}.
Note that the perfect ring (\ie a ring with no attached leads)
supports eigenstates of $q$ for integer $q$'s corresponding to
energy eigenvalues $E_n=n^2h^2/2\mu l^2$ (corresponding to $q=n$),
where $\mu$ is the effective mass of the electron.

The typical behavior of $T$ versus $\theta$ (for arbitrary $q$)
exhibiting transmission zeros for half odd integer values of
$\theta$ and local minima for integer values of $\theta$ is shown in
the thin line curve of Fig.\ref{transones}. However, for special
values of $q$ in the proximity of integer $q$ values, sharp
transmission ones' appear (thick line in Fig.\ref{transones}).

The flux periodicity of $1$ reflected in the pattern of
Fig.\ref{transones} may give the impression that the phenomenon of
electron transmission in the ABI is also flux periodic with the
{\em same} period $1$. However, this expectation is {\em not}
borne out by the following results relating to the TA. For
$q=n+0.01$ ($n$ being an integer), \ie for $q$-values very close
to integer values $n$ corresponding to the transmission ones, the
trajectory in the complex $t$-plane is shown in Fig.\ref{scurve1}.
\begin{figure}[h]
\vskip 2cm
  \includegraphics[width=2in]{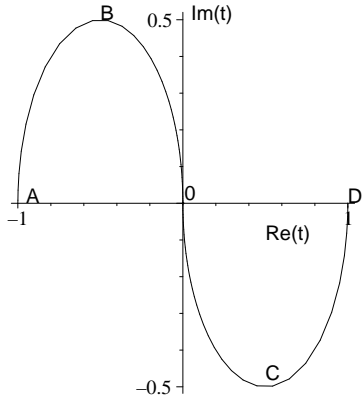}  
\caption{\label{scurve1}Trajectory generated in the complex $t$-plane as the flux is
  varied. The case considered is that of a symmetric ABI.}
\end{figure}
\noindent
It starts from a point $A$
for $\theta=0$, covers the path $ABO$ as $\theta$ is made to vary from
$0$ to $1/2$ and thereafter follows the path $OCD$ as $\theta$ varies
from $1/2$ to $1$, reaching the point $D$. If $\theta$ is varied
further from $1$ to $2$, the path $DCOBA$ is followed, thus retracing
the path to generate a periodic trajectory. Thus {\em two} flux
periods are necessary to complete one periodic cycle in the
$t$-plane. The above description holds also for $q=n-0.01$; only the
shape of the trajectory appears reflected about the real axis. We note
that the above two values of $q$ corresponds to energy values--one
slightly {\em above} and the other slightly {\em below} the discrete
eigenvalues of the electron energy (in the perfect ring).

For $q$ values covering the range $n\le q\le n+\frac{1}{2}$, the
shape of the trajectory remains unaltered; only the curve gets
rotated progressively in the anticlockwise (or clockwise, depending
on the direction of the flux) direction to attain the maximum
rotation  by an angle $\pi /2$. As $q$ is increased further to cover
the range $n+\frac{1}{2} \le q\le n+1$, the trajectory has a
``flipping point'' when it's shape gets completely reflected about
the Im($t$) axis as $q$ crosses the value $n+\frac{1}{2}$, but keeps
rotating in the same direction by a further amount of $\pi /2$. The
final shape is {\em not} back to its initial shape. Further
variation of $q$ from $n+1$ to $n+2$ is required to generate one
complete periodic cycle. This behavior is depicted in
Fig.\ref{srotating3}.
\begin{figure}
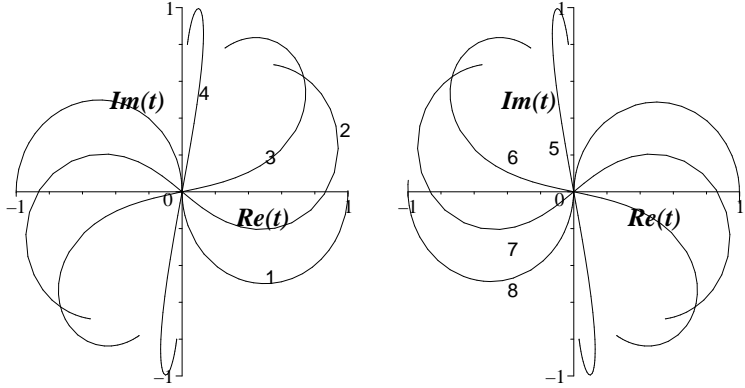
 
\vskip 1cm
\includegraphics[width=2in]{srotating1.eps}
\includegraphics[width=2in]{srotating3.eps}
\caption{\label{srotating3}Complex $t$-plane plots as the flux is
  varied. Curves marked 1,2,3 and 4 correspond to $q$ values
  between $0$ and $1/2$. The ones marked 5,6,7 and 8 correspond
  to $q$ values between $1/2$ and $1$.}
 \end{figure}
\subsection{The asymmetric ring interferometer}
In order to fix the ideas, we first consider the specific example
of a ring with $l_1:l_2=2:3$. The cases corresponding to other
rational fraction ratios are similar. The first interesting
observation is that the trajectories form closed curves only upon
varying $\theta$ over five ($=2+3$, the sum of the numbers in the
ratio $l_1:l_2$) flux periods. A typical plot is presented in
Fig.\ref{partpetal} for $q=1/4$ (\ie $k=\pi /2l$) when the flux is
varied in the range $0\le\theta\le 1$. Note that the shape of this
curve is similar to the curve in Fig.\ref{scurve1}, but the curve
does {\it not} pass through the origin. In the case of the
symmetric ABI, further variation beyond the  above range only
results in the repetition of this trajectory.
\begin{figure}
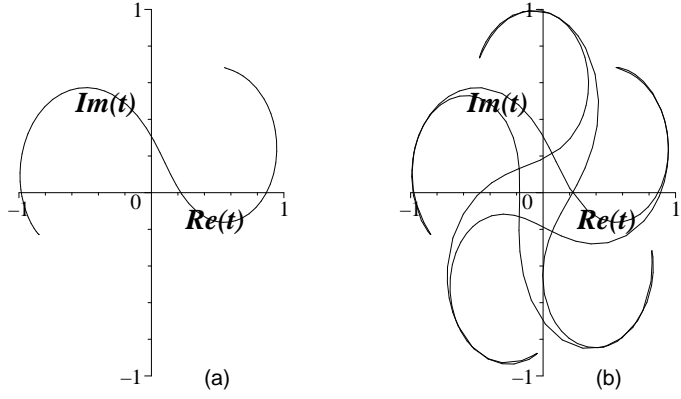
  
\vskip 1cm
  \includegraphics[width=2in]{part_petal.eps}
  \includegraphics[width=2in]{petals_piby2.eps}
\caption{\label{partpetal}Complex $t$-plane plot at $q=1/4$ for
  $l_2:l_3=2:3$ as the flux is varied over (a)one flux period, (b)five
  flux periods.}
\end{figure}
\noindent
 Hut in the present case, further variation of flux leads
to evolution along different directions and when the variation
covers five flux-periods (\ie , $0\le\theta\le 5$, a closed
flower-like pattern results (see Fig.\ref{petals}). Further
variation of flux leads to reparation of this pattern. This behavior
is indicative of an over-all periodicity of five in the underlying
phenomenon. Note that for this value of $q$, the $T$ versus $\theta$
plot does {\it not} reveal this $5$-fold periodicity (see
Fig.\ref{trans_petals}).
\begin{figure}[h]                          
\vskip 2cm
  \includegraphics[width=2in]{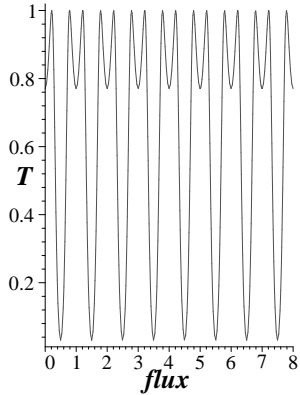}
\caption{\label{trans_petals}Variation of $T$ with $\theta$ for $q=1/4$.}
\end{figure}
Some plots for a few more typical $q$ values are shown in
Figs.{\ref{petals}.
\begin{figure}
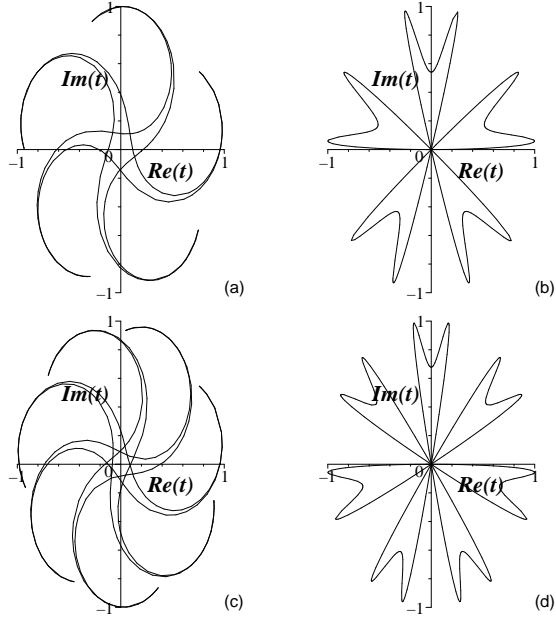
                              
\vskip 2cm
  \includegraphics[width=4.0cm]{five_piby3.eps}
  \includegraphics[width=4.0cm]{five_2pi.eps}\\[1cm]
  \includegraphics[width=4.0cm]{seven_piby3.eps}
  \includegraphics[width=4.0cm]{seven_4pi.eps}
\caption{\label{petals}Typical complex $t$-plane plots. The parameters
  chosen are: (a)$l_2=2/5$, $l_3=3/5$, $q=1/6$, $0\le\theta \le 5$;
  (b)$l_2=2/5$, $l_3=3/5$, $q=1$, $0\le\theta \le 5$; (c)$l_2=3/7$,
  $l_3=4/7$, $q=1/6$, $0\le\theta \le 7$; (d)$l_2=3/7$,
  $l_3=4/7$, $q=2$, $0\le\theta \le 7$.}
\end{figure}
Note that the passage of the trajectory through the origin signifies
the presence of a transmission zero.

When the value of the ratio $l_2:l_3$ is not a rational fraction, the
pattern generated is {\it not} a closed curve. Figure\ref{flower}
presents an example of such a pattern for $l_2=1/\sqrt{5}$ and
  $l_3=1-1/\sqrt{5}$. In this case the
trajectory {\it never closes} and as $\theta$ is varied over longer
and longer intervals, the pattern generated turns out progressively
denser--the behavior reflected in the figure.
\begin{figure}
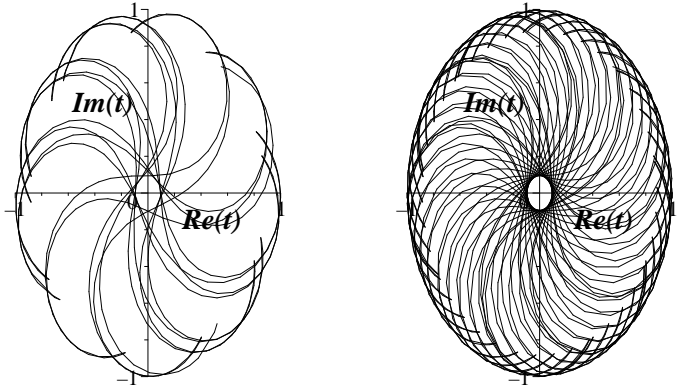
                                 
\vskip 2cm
  \includegraphics[width=2in]{flower.eps}
  \includegraphics[width=2in]{flower1.eps}
\caption{\label{flower}Complex $t$-plane plot for $l_2=1/\sqrt{5}$,
  $l_3=1-1/\sqrt{5}$. The plot on the left is drawn for $0\le\theta\le
  15$ and the one on the right, for $0\le\theta\le 45$.}
\end{figure}
Clearly, the phenomenon is {\it not} flux-periodic. We now report on
the behavior of the complex $t$-plots when, in stead of the flux,
the momentum $q$ is made to vary.

\section{Type II $t$-plots}
The complex plane plot of $t$ as the electron momentum is varied have
been discussed in the literature earlier\cite{kim_cho_kim_ryu} for the
case of an ABI with an embedded quantum dot and also for an embedded
double-barrier. We are not aware of such a calculation for a bare ABI.

For the symmetric ring ABI, we obtain the typical behaviors for
different cases corresponding to fixed typical values of the flux.
Figure \ref{type2sym} displays these behaviors.
\begin{figure}
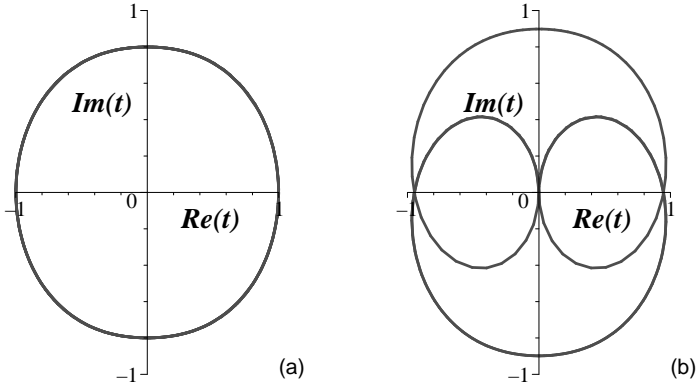

\vskip 2cm                                 
\includegraphics[width=2in]{zeroflux.eps}
\includegraphics[width=2in]{highflux.eps}
\caption{\label{type2sym}Type II $t$-plots for the symmetric ring
  ABI. (a)$\theta =0$, (b) $\theta =5$.}
\end{figure}
Note that when the flux is zero (case (a)), the closed curve is an
ellipse and it does {\it not} pass through the origin indicating the
absence of transmission zeros. However, the presence of even a tiny
flux causes transmission zeros to appear; thus the plots (cases b,c
and d) exhibit curves passing through the origin. This qualitative
difference between the two cases is remarkable. We note in passing
that increase of flux to higher magnitudes does not alter the
appearance shown in Fig.\ref{type2sym}b.

The case of an asymmetric ABI with the ratio $l_2:l_3$ being a
rational fraction is shown in Figs.\ref{type2asym}.
\begin{figure}
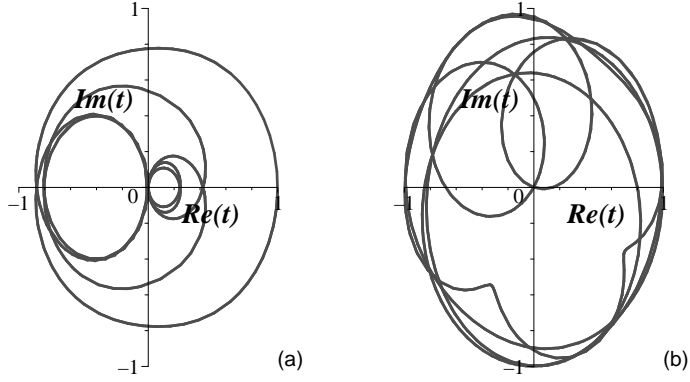
                                  
\includegraphics[width=5cm]{zeroflux_asym.eps}
\includegraphics[width=5cm]{highflux_asym.eps}
\caption{\label{type2asym}Type II $t$-plots for an asymmetric ring
  ABI. (a)$\theta =0$, (b) $\theta =5$, and $0\le q\le 10$.
  In both cases, $l_2=2/5$ and $l_3=3/5$.}
\end{figure}
The variation of $q$ over larger and larger intervals does
{\it not} generate additional features. In fact, the plot for $0\le
q\le 50$ is no different from that presented in Fig.\ref{type2asym}b. This implies a kind of
(rather complicated) $q$-periodicity in the phenomenon.

When the ratio $l_2:l_3$ is irrational, the type II $t$-plots, just as
the type I $t$-plots, do not exhibit closed
curves. Figure\ref{type2irr} illustrates this point.
\begin{figure}
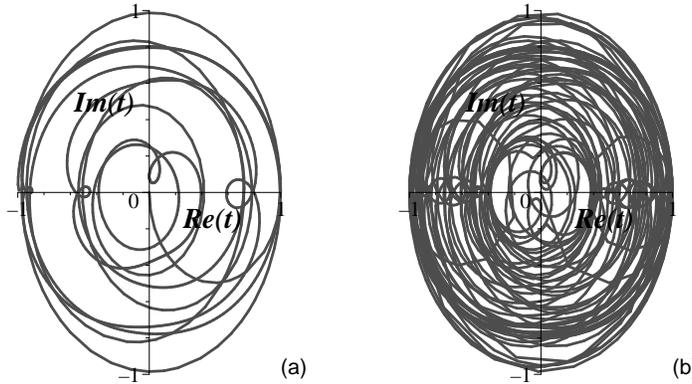

\vskip 1cm                                   
\includegraphics[width=2in]{lowflux_irr.eps}
\includegraphics[width=2in]{lowflux_irr1.eps}
\caption{\label{type2irr}Type II $t$-plots for an ABI with
  $l_2=1/\sqrt{5}$ and $l_2=1-1/\sqrt{5}$. (a) $\theta =5$ and $0\le
  q\le 10$, (b)  $\theta =5$ and $0\le q\le 50$.}
\end{figure}
Note that as the variation of $q$ spreads over larger intervals, the
pattern generated becomes denser.
With these results we end with our conclusions.

\section{conclusion}
Our results establish the utility of the type I $t$-plots in the
characterization of the ABI. The information obtained via both types
of plots is very rich and perhaps need to be pursued further. The
fact that the flux periodicity nature of the transmission
coefficient $T$ does not imply corresponding periodicity in the
underlying phenomenon is clearly brought out in the present
calculations. It may be of interest to study the changes arising in
the plots when an impurity is embedded in one of the arms of the
interferometer. We have obtained a host of new results by way of
constant transmission contour plots and also investigated the effect
of electric field acting on one of the arms of the interferometer.
These results will be reported later.

\section*{Acknowledgements}
MVAK wishes to thank Prof. K. Venkataramaniah for his
encouragement. DS wishes to thank the Director, Institute of
Physics for his kind hospitality.

\end{document}